\documentclass[prb,preprint]{revtex4-1} 

\usepackage{amsmath}  
\usepackage{amsfonts} 
\usepackage{graphicx} 

\begin{document}


\title{Helicity and nuclear $\beta$ decay correlations}

\author{Ran Hong}
\email{hongran@uw.edu} 
\altaffiliation[current address: ]{Argonne National Laboratory, 9700 S Cass Ave, Building 362, 
  Argonne, IL 60439, USA}
\author{Matthew G. Sternberg}
\author{Alejandro Garcia}
\affiliation{Physics Department and CENPA, University of Washington, Seattle, Washington 98195}



\date{\today}

\begin{abstract}
We present simple derivations of nuclear $\beta$-decay correlations with an emphasis on the special role of helicity. This provides a good opportunity to teach students about helicity and chirality in particle physics through exercises using simple aspects of quantum mechanics. In addition, this paper serves as an introduction to nuclear $\beta$-decay correlations from both a theoretical and experimental vantage. This article can be used to introduce students to ongoing experiments searching for hints of new physics in the low-energy {\em precision} frontier. 
\end{abstract}

\maketitle 

\section{Introduction}
\label{sec:introduction}
Helicity is the projection of the spin of a particle onto the direction of its momentum. Helicity plays an important role in modern physics and a good understanding of the associated rules is important for interpreting many atomic, nuclear, and particle physics experiments. Thus, development of intuition with respect to different aspects of helicity in quantum mechanics is a worthwhile exercise for classes that are taught to advanced undergraduate or beginning graduate students.

In this paper we concentrate on the angular correlations (for example, between the spin of parent nucleus and the direction of emitted $\beta$ particle) that arise in nuclear $\beta$ decays due to the combination of conservation of angular momentum and the helicity of the leptons. Because we concentrate on nuclear $\beta$ decays we will use ``weak interaction'' as a synonym of ``{\em charged} weak interactions''. Measurements of correlations from nuclear $\beta$ decay originally established the vector minus axial-vector nature of the weak currents, known as $V-A$, about 50 years ago. We will explain the significance of $V-A$ and describe new measurements of decay correlations being pursued in search of new physics with helicity properties that differ from the prescriptions of the standard model of particle physics (SM). 

The correlations can be calculated using trace techniques of Dirac's $\gamma$ matrices and are sometimes brought up in this context as exercises for students learning relativistic quantum mechanics or field theory. On the other hand, the calculated expressions are often presented to students without this training along with comments to show their plausibility. In contrast, we present a more accessible derivation using tools learned in the elementary quantum mechanics classes for which most advanced undergraduates should be well equipped. Moreover, those students capable of doing the calculations via trace techniques may not appreciate that the correlations arise simply from the conservation of angular momentum and the left-handedness of the emitted leptons. It is often far too easy to let the mathematical formalism overshadow the elegant and beautiful physical principles at work.

The present paper is intended for a broad audience; it is well suited for students that have completed an introductory quantum mechanics course, while providing supplementary material for the more advanced readers with experience in quantum field theory. We begin with a simple derivation of the $\beta$ asymmetry with respect to the polarization of the parent nucleus in Sect.~\ref{sec:helicity}. In Sect.~\ref{sec:currents} we will present a brief description of the weak-interaction Hamiltonian. Although we present here the interactions using Dirac's $\gamma$ matrices, we think all that is needed to follow that section is a brief introduction to the Dirac equation and instructors can use Appendix~\ref{sec:appendix1} as a guideline. In Sect.~\ref{sec:Fierz} and \ref{sec:enu correlation} we derive the so-called {\em Fierz interference} term and the $e-\nu$ correlation.\footnote{In this paper, we use ``$e-\nu$ correlation'' to represent both electron-antineutrino and positron-neutrino correlations.} The derivations are presented alongside a brief historical narrative, including the story of several experiments which wrongly led physicists to an incorrect theory of the weak interactions. A few contemporary experiments involving $\beta$-decay correlations are introduced in Section~\ref{sec:ContemporaryExp}.

\section{Helicity and chirality properties of the weak interaction}
\label{sec:helicity}
\begin{figure}[ht]
\includegraphics[width=1.6in]{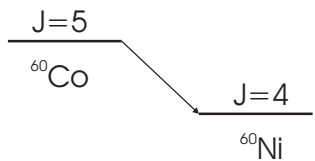}
\caption{Decay scheme for $^{60}{\rm Co}$.}
\label{fig:60Co-scheme}
\end{figure}

\begin{figure}[ht]
\includegraphics[width=1.6in]{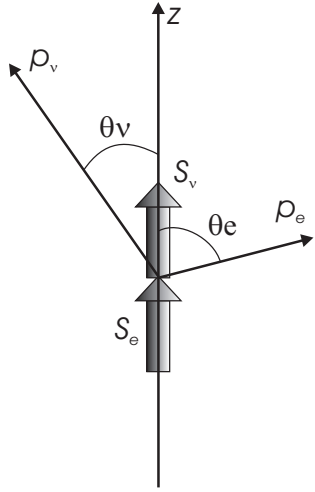}
\caption{Correlation between the initial polarization and the electron direction: in the decay of $^{60}{\rm Co}$ one unit of angular momentum is lost by the nucleus and has to be carried out by the leptons. Since the latter are spin-1/2 objects they have to both align their spins in the direction of the initial polarization. This, in conjunction with the helicities of the particles, determines the emission probabilities.}
\label{fig:SpinCorrelation}
\end{figure}

The main features of $\beta$ decay are described in many textbooks.\cite{ko:66,kr:87,he:07} Here we briefly discuss the aspects that are relevant for the present discussion. We start by considering the correlations between the spin polarization of the parent nucleus and the direction of the emitted electrons in the famous experiment of Wu et al.~\cite{wu:57} which was one of the first experiments to confirm the hypothesis of parity violation by the weak interactions put forward by Lee and Yang.\cite{leeyang:56} Wu and collaborators polarized a sample of radioactive $^{60}{\rm Co}$ atoms and observed the distribution of emitted electrons relative to the direction of the initial nuclear spin polarization. The corresponding decay scheme is shown in Fig.~\ref{fig:60Co-scheme}. 
In this section, we first consider the transition from the $M=5$ initial state ($M$ is the ``magnetic quantum number'') to the $M=4$ final state. Following the $\beta$-decay transition one unit of angular momentum along the direction of the initial polarization is lost from the nucleus, and this angular momentum must be carried away by the lepton spins.\footnote{In this paper, we focus only on ``allowed decays'' in which the leptons (the $\beta$ and $\bar \nu$) carry no orbital angular momentum so only spins are of concern.} Thus, the spin projections of the two leptons onto the $z$ axis have to be $+1/2$ each, as shown in Fig.~\ref{fig:SpinCorrelation}.

For an electron (or any other spin-1/2 particle) emitted at an angle $\theta$ relative to the $z$-axis, the positive or negative helicity states $|\theta_{\pm}\rangle$ can be expressed as linear combinations of spin-up and spin-down states along the $z$-axis $|\pm\rangle$:\cite{Sakurai}$^{,}$\footnote{This formula can also be derived by considering a spin-1/2 particle moving through two Stern-Gerlach devices: the first one oriented along the direction of polarization (corresponding to the angular momentum defined by the nuclei) and the second one in a random polar direction (corresponding to the emission of the left-handed electron in a particular direction)}
\begin{align}
|\theta_+\rangle&=\cos{(\theta/2)}  |+\rangle- \sin{(\theta/2)} |- \rangle \nonumber\\
|\theta_-\rangle&=\sin{(\theta/2)}  |+\rangle+ \cos{(\theta/2)} |- \rangle.
\label{Rot_Mat}
\end{align}
For electrons in the negative-helicity state $|\theta_-\rangle$, the probability of finding the $+1/2$ spin projection onto the $z$ axis is $\sin^2{(\theta/2)}$. Thus, the probability per solid angle $dP/d\Omega$ of emitting such an electron is given by:
\begin{eqnarray}
\frac{dP}{d\Omega}=\frac{\sin^2{(\theta/2)}}{2\pi} = \frac{(1-\cos{\theta})}{4\pi}. 
\label{eq:helicity01}
\end{eqnarray}
On the other hand, for positive-helicity electrons 
\begin{eqnarray}
\frac{dP}{d\Omega}=\frac{\cos^2{(\theta/2)}}{2\pi} = \frac{(1+\cos{\theta})}{4\pi}. 
\label{eq:positive_helicity_distribution}
\end{eqnarray}

The parity transformation (which inverts the sign of coordinates) turns a negative-helicity particle into a positive-helicity particle. Under the assumption of parity conservation the emitted leptons from $^{60}{\rm Co}$ should show no preference for either helicity state. The sum of the distributions in Eq.~(\ref{eq:helicity01}) and Eq.~(\ref{eq:positive_helicity_distribution}) with equal weights does not depend on $\theta$. Thus, one would expect the electrons to be emitted uniformly in all directions. In one of the greatest surprises in modern physics Wu and others\cite{wu:57, garwin:57, friedman:57} found quite the opposite to be true. They discovered that the electron angular distribution was close to Eq.~(\ref{eq:helicity01}). It appeared as though only negative helicity electrons were being emitted. Much to the surprise of researchers at that time, and still a surprise to many of us today, the laws of nature appear to have a preferred handedness.

In order to understand how this fact has been implemented into the Standard Model we need a new concept: {\em chirality}. 
Note that for a massive particle the helicity of a particle is not a Lorentz invariant which does not change under rotations or boosts of the reference frame. An observer moving faster than the particle will see its helicity in the opposite direction. By contrast, chirality is a Lorentz invariant. We follow Konopinski\cite{ko:66} who gives an intuitive description of chirality. Consider an electron moving in the $+z$ direction with momentum $p$, energy $E$, and spin along the $+z$ direction. If we want to measure the velocity of this electron, we need to take an infinitesimal time period, so the energy uncertainty goes to infinity according to Heisenberg's uncertainty principle. However, particles with infinite energy move at the speed of light, $c$, so the results of such a measurement are $\pm c$. We call this {\em internal velocity}. The physical state moving at velocity\footnote{In this paper we use natural units so that $c=1$ and $\hbar=1$.} $v=p/E$ can be described 
as a combination of a ``forward-motion'' along the path at speed $c$ with a probability $(1+p/E)/2$ and a ``backward-motion'' at speed $c$ with a probability $(1-p/E)/2$. The physical velocity $v$ is just the mean velocity of this motion. We use $u_{\uparrow\downarrow}(E,{\bm p})$ to represent physical states with definite momentum, energy and spin (up or down) along the $z$ direction, and $\phi_{\uparrow\downarrow}(\pm c)$ to represent the states with internal velocity $+c$ or $-c$ and spin up or down along the $z$ direction. Then the physical state $u_{\uparrow}(E,{\bm p})$ can be expressed as a linear combination of $\phi_{\uparrow}(+ c)$ and  $\phi_{\uparrow}(- c)$:
\begin{eqnarray}
\label{eq:chirality_decompose}
u_{\uparrow}(E,{\bm p})=\sqrt{\frac{1+p/E}{2}}\phi_{\uparrow}(+c)+\sqrt{\frac{1-p/E}{2}}\phi_{\uparrow}(-c).\nonumber\\
\end{eqnarray} 
A particle's chirality can be defined as ``the spin projection onto its internal velocity direction''. In this sense, the state $\phi_{\uparrow}(+c)$ has right-handed chirality and $\phi_{\uparrow}(-c)$ has left-handed chirality. Therefore, a state with well-defined helicity, momentum and energy, like $u_{\uparrow}(E,{\bm p})$, is a linear combination of two states with opposite chirality and relative amplitudes as in Eq.~(\ref{eq:chirality_decompose}). If we build up a state like $\phi_{\uparrow}(+c)+\phi_{\downarrow}(-c)$ which has definite right-handed chirality, it is not a free-particle state because its spin projection is not $\pm1/2\hbar$. If the particle is massless, then only one internal velocity state describes it, so the free-particle state contains only one chirality component. In this case helicity and chirality describe the same property of the particle. Left-handed chirality is equivalent to negative helicity for massless particles. 

A formal description of free fermions is shown in Appendix~\ref{sec:appendix1} using Dirac spinors. The motion of a free fermion is governed by the Dirac equation, which has two positive energy solutions, i.e. particle solutions, with well-defined energies, momenta and helicities (see Eq.~(\ref{eq:Dirac Solutions})). This is due to the fact that both momentum and helicity operators commute with the free particle Hamiltonian. The spinor parts of these two solutions $u_{\uparrow\downarrow}(E,{\bm{p}})$ have opposite helicities, but both of them have non-zero left-handed chirality projections with amplitudes
\begin{eqnarray}
|P_{L}u_{\uparrow}(E,{\bm{p}})|=\sqrt{1-p/E}/\sqrt{2},\nonumber\\
|P_{L}u_{\downarrow}(E,{\bm{p}})|=\sqrt{1+p/E}/\sqrt{2},
\label{eq:chirality_projection}
\end{eqnarray}
where $P_{L}$ is the left-handed chirality projection operator. This is consistent with Eq.~(\ref{eq:chirality_decompose}). 

In the SM weak interactions involve only particles with left-handed chirality. Therefore, for massive particles like electrons, both helicity states are involved in the weak interaction with amplitudes expressed in Eq.~(\ref{eq:chirality_projection}), and Eq.~(\ref{eq:helicity01}) for $dW/d\Omega$ has to be modified to take this into account. The correct expression for $dW/d\Omega$ (for ``$M\to M-1$'' transitions) in the SM is
\begin{alignat}{1}
\frac{dW}{d\Omega}&=\frac{
\sin^2{(\theta/2)}}{2\pi} 
\frac{
\left(
1+p_e/E_e 
\right)}
{2} + 
\frac{\cos^2{(\theta/2)}}{2\pi} 
\frac{\left(
1-p_e/E_e 
\right)}{2}\nonumber\\
&=\frac{\left(
1- (p_e/E_e) \cos{\theta}
\right)}{4\pi}.
\label{eq:correlation01}
\end{alignat}

In general, given that the nuclei experience a change of projection of angular momentum $\Delta M$ (defined as $M_{parent}-M_{daughter}$) along a quantization direction $z$, the electron will have an angular distribution:
\begin{eqnarray}
\frac{dW}{d\Omega}=\frac{1 - {\Delta M} (p_{e}/E_e)\cos{\theta}}{4\pi}
\label{angular_distr_M}
\end{eqnarray}
with $\Delta M=0, \pm 1$. If the $^{60}{\rm Co}$ nucleus is in a state with $M \neq \pm 5$ (for example $M=3$), the final state can have different $M$ values ($M=2,3,4$). Usually the spin projection of the final state of the daughter nucleus is difficult to detect, so one may sum over the probabilities of final states with different $M$. The probabilities of decaying into these final states are not equal, but proportional to the square of the Clebsch-Gordan coefficients $\langle J_f,M_f;J_l,M_l|J_i,M_i\rangle$, where $J_i,M_i$ are the spin and spin projection of the initial nuclear state, $J_f,M_f$ are the spin and spin projection of the final nuclear state, and $J_l=1$, $M_l=0,\pm1$ are the total angular momentum and its projection taken away by the leptons. 
After summing over the final nuclear states, one gets the angular distribution:
\begin{eqnarray}
\frac{dW}{d\Omega}=\frac{1 - \frac{M_i}{J_i}(p_{e}/E_e)\cos{\theta}}{4\pi}.
\label{m/j}
\end{eqnarray}
The proof of this equation is a bit too long to reproduce here, but it can be a good exercise for students when learning angular momentum raising and lowering operators and Clebsch-Gordan coefficients. Defining the polarization vector for the initial ensemble of nuclei as ${\bm P}=\langle{\bm J_i}\rangle/J_i$, it follows that
\begin{eqnarray}
\frac{dW}{d\Omega}=1 + A {\bm P} \cdot \frac{\bm p_e}{E_e}
\end{eqnarray}
with the $\beta$ asymmetry correlation coefficient\footnote{This calculation is only valid for the case of $J \rightarrow J-1$ decays such as the decay of $^{60}$Co, because Eq.~(\ref{m/j}) does not apply for other cases. More advanced readers can try to derive the $\beta$ asymmetry coefficient $A$ for all initial and final $J$'s, and the results can be found in Ref.~\cite{ja:57}.} $A=-1$ for the decay of $^{60}{\rm Co}$.

Note that the expressions above show that the decay rate varies under the parity transformation which flips the sign of ${\bm p_{e}}$ but not ${\bm P}$. The experimental determination of $A$ by Wu et al.\cite{wu:57} showed clearly that parity conservation was violated by the weak interactions. Another observable worth discussing is the polarization of emitted electrons, $P_e$. Using Eq.~(\ref{eq:chirality_projection}), one can derive
\begin{eqnarray}
P_e = \frac{(1-p/E)(+1)+(1+p/E)(-1)}{(1-p/E)+(1+p/E)}=-p/E.
\end{eqnarray}
This was later directly confirmed by several experiments.\cite{Frauenfelder:57,Frauenfelder:57-2,Klinken:66}

We have shown that assuming electrons emitted in $\beta$ decay have left-handed chirality leads to good agreement with experiment, but this does not completely determine the formalism of the weak interaction. One important missing piece of the theory is the helicity of the antineutrino, which is almost impossible to measure directly. The relationship between chirality and helicity for {\em antiparticles} is discussed in Appendix~\ref{sec:appendix1}.  According to the SM the chirality of antineutrinos involved in weak interactions is also left-handed, and left-handed antiparticles have {\em positive} helicity. If we take the SM description of antineutrinos for granted, namely that they have positive helicity, the angular distribution of the antineutrino around the direction of nuclear polarization is:
\begin{eqnarray}
\frac{dW}{d\Omega}=1 + B {\bm P} \cdot \frac{\bm p_{\bar \nu}}{E_{\bar \nu}}.
\end{eqnarray}
with the antineutrino asymmetry correlation coefficient $B=+1$ for the decay of $^{60}{\rm Co}$. As we shall see in the following sections, neutrino helicities were determined through indirect measurements and the complete weak interaction formalism was built in the 1960s.

\section{Scalar, Vector, and Tensor Currents}
\label{sec:currents}
To understand the Fierz interference and the $e-\nu$ correlation described in the following two sections, one has to go one level deeper into the weak interaction theory and understand the formalism of its Hamiltonian. At the time when Lee and Yang proposed that parity was violated, little was known about the weak interaction. Dirac had already shown how to solve problems involving electromagnetic interactions within a quantum theory that correctly takes into account relativity.\cite{Dirac1} For example, for electron scattering from a proton at low-momentum transfer (so that internal nucleon excitations can be neglected) the interaction can be expressed as a product of a nuclear current, a {\em propagator} for the photon, and an electronic current:\cite{ha:84,gr:08,ca:13} 
\begin{eqnarray}
H_{EM}= (\bar\psi_p \gamma^\mu \psi_p) \frac{-e^2}{(-q^2)} (\bar\psi_e \gamma_\mu \psi_e).
\end{eqnarray}
Here the $\psi's$ are Dirac spinor operators which can annihilate a particle (with certain momentum, energy, etc.) in the initial state or create an antiparticle from vacuum, $\bar \psi$ means $\psi^{\dagger}\gamma^0$, and $q^2$ is the square of the 4-momentum transfer in the scattering process. The $\gamma^\mu$'s are Dirac's $\gamma$ matrices with the properties:
\begin{eqnarray}
\gamma^\mu \gamma^\mu &=& 
\left\{
\begin{array}{l}
~~I~{\rm for}~\mu=0\\
-I~{\rm for}~\mu=1,2,3
\end{array}
\right.
\nonumber\\
\gamma^\mu \; \gamma^\nu &=& - \gamma^\nu \; \gamma^\mu~~{\rm for~\mu \ne \nu}; \mu,\nu=0,1,2,3,
\label{eq:commutation} 
\end{eqnarray}
where $I$ stands for a $4\times4$ identity matrix. To follow the derivations in the rest of this paper, it is adequate to remember the properties of the $\gamma$ matrices in Eq.~(\ref{eq:commutation}) without mastering the full expressions of them shown in Appendix~\ref{sec:appendix1}.
Fermi proposed a similar structure for the weak interactions:\cite{Fermi1934,WilsonAjp1968}
\begin{eqnarray}
H_{Fermi}=\frac{G}{\sqrt{2}} (\bar\psi_p \gamma^\mu \psi_n) \; (\bar\psi_e \gamma_\mu \psi_\nu)+h.c.,
\end{eqnarray}
where $h.c.$ indicates the hermitian conjugate. The first term in $H_{Fermi}$ describes an incoming neutrino (or an outgoing antineutrino) and an outgoing electron, while the $h.c.$ term describes an incoming electron (or an outgoing positron) and an outgoing neutrino. For massive bosonic force carriers the propagator denominator is $M^2-q^2$, where $M$ is the mass of the carrier. Because the mass of the carrier for the weak interactions is much larger than the momenta involved in the nuclear weak transitions the propagator is constant to a very good approximation. When Fermi proposed this form for the interaction it was unknown what the mass of the carrier was, but nowadays we know $M_W\sim80$ GeV while the momentum transfer in nuclear $\beta$ decay is of the order of a few MeV. In fact, the {\em weakness} of the weak interaction in nuclei is due to the large mass of the W compared to the energy released in $\beta$ decays. Although Fermi focused on the hypothesis of the vector interaction he indicated that other possibilities were allowed. Many years later Lee and Yang\cite{leeyang:56} explicitly included parity violation. Thus, the general Hamiltonian for the weak interaction can be expressed as:
\begin{eqnarray}
H_{int}=\sum_{i=S,P,V,A,T} (\bar\psi_p O^i \psi_n)\;
(C_i \bar\psi_e O_i \psi_\nu+\nonumber\\
 C_i^\prime \bar\psi_e O_i \gamma^5 \psi_\nu)+h.c.
\label{eq:LYHamiltonian}
\end{eqnarray}
The $C_i$'s are constants that could be determined experimentally and the operators $O_i$'s are
\begin{eqnarray}
O_S &=& 1 \nonumber\\
O_P &=& \gamma^5\nonumber\\
O_V &=& \gamma_\mu \label{eq:spvat} \\
O_A &=& i \gamma_\mu \gamma^5 \nonumber\\
O_T &=& \sigma_{\mu \nu} /\sqrt{2}=-i \left(\gamma_\mu \gamma_\nu - \gamma_\nu \gamma_\mu \right)/(2\sqrt{2}).\nonumber
\end{eqnarray}
The corresponding currents $\bar\psi O^i \psi$ are called, respectively, scalar, pseudo-scalar, vector, axial-vector (or pseudo-vector), and tensor. The additional gamma matrix included here, $\gamma^5$, can be expressed in terms of the other four, $\gamma^5=i \gamma^1 \gamma^2 \gamma^3\gamma^0$. Using Eq.~(\ref{eq:commutation}) one can show that
\begin{eqnarray}
\gamma^5 \; \gamma^5 = I
~{\rm and}~
\gamma^\mu \; \gamma^5 = - \gamma^5  \; \gamma^\mu
\label{eq:commutation2}
\end{eqnarray}
The property of the operators under parity transformation becomes evident if one considers what happens when the coordinates are inverted. Note that under the parity transformation the spatial components of $p_\mu$ are odd and its time-like component (energy) is even. The kinetic term in the Dirac Equation $\gamma^\mu p_\mu$ is a Lorentz scalar\footnote{Because the Dirac Equation reconciles quantum mechanics and special relativity, its form is invariant under Lorentz transformation. Strictly speaking, it is the term $\gamma^\mu p_\mu \psi$ that is a Lorentz scalar.} (or Lorentz invariant). Because a Lorentz scalar has even parity, a Lorentz scalar can only be the inner product of two vectors with the same parity. Therefore, the spatial components of $\gamma^\mu$ should be parity-odd, while its time-like component should be parity-even. Because the $\gamma^5$ matrix is the product of one time-like matrix and 3 space-like ones, it is parity-odd, and multiplication by it reverses the parity property of all operators. This $\gamma^5$ matrix is by definition the chirality operator.\cite{ha:84} Consequently, the operators:
\begin{eqnarray}
P_{L/R}=(1 \pm \gamma^5)/2
\label{projectors}
\end{eqnarray}
are the projectors onto left- and right-handed chirality states. Under parity transformation, $P_{L/R}$ turns into $P_{R/L}$, so a left-handed state transforms to a right-handed state. In the relativistic limit, when the masses of the particles are negligible compared to their energies, chirality is equivalent to helicity, so $\gamma^5$ becomes the helicity operator and the two projectors above become the projectors onto the helicity states. The interaction Hamiltonian in Eq.~(\ref{eq:LYHamiltonian}) can be re-written in terms of left- and right-handed lepton spinors. For the vector and axial-vector currents:
\begin{eqnarray}
\label{eq:lrva}
H^{VA}_{int}=\sum_{i=V,A}
(\bar\psi_p O^i \psi_n)\;
\left(
(C_i+C_i^\prime)\; \bar\psi_e^L O_i \psi_\nu^L+ \right. \\
\left.
(C_i- C_i^\prime)\; \bar\psi_e^R  O_i \psi_\nu^R
\right).\nonumber
\end{eqnarray}
For the scalar and tensor currents (potentially new physics):
\begin{eqnarray}
\label{eq:lrst}
H^{ST}_{int}=\sum_{i=S,T}
(\bar\psi_p O^i \psi_n)\;
\left(
(C_i+C_i^\prime)\; \bar\psi_e^R O_i \psi_\nu^L+ \right.  \\
\left.
(C_i- C_i^\prime)\; \bar\psi_e^L O_i \psi_\nu^R
\right). \nonumber
\end{eqnarray}
The notation $\psi^{L/R} =P_{L/R}  \; \psi$ is used in Eq.~(\ref{eq:lrva}) and Eq.~(\ref{eq:lrst}). We have ignored the pseudo-scalar currents because they turn out to be very small in nuclear $\beta$ decays. In this article we will assume the constants $C_i$ to be real. As we will describe later, present limits on these non-Standard Model couplings ($C_S,C_S^\prime,C_T,C_T^\prime$) are of order $10\%$.~\cite{severijns:06} Allowing for complex phases brings in time-reversal symmetry violation, which is very interesting,\cite{emit} but subject for another paper. Notice that while the vector and axial-vector currents couple incoming and outgoing particles with identical chiralities, the scalar and tensor currents do the opposite. This is a direct consequence of Eq.~(\ref{eq:commutation2}). 

Though Wu et al. found that the electrons from nuclear $\beta$ decays mostly have negative helicity and thus only electrons with left-handed chirality are involved in nuclear $\beta$ decays, they could not determine the helicity of the emitted antineutrino. Additional experiments were proposed to determine whether the currents were scalar, vector, axial-vector or tensor, or some combination of these. As we will see, eventually they determined that the weak interaction is primarily mediated by vector and axial-vector currents.

The form of hadronic currents also affects the changes of nuclear angular momenta in $\beta$ decays. Conventionally, nuclear $\beta$ decays are  classified according to the change in angular momentum $J$ and isospin $T$. Some basics on isospin are given in Appendix~\ref{sec:appendix3}.  ``Fermi transitions'' (F) are those with $\Delta J = 0$, $\Delta T = 0$ and ``Gamow-Teller transitions'' \cite{ga:36}(GT) have $\Delta J = \pm1,0$, $\Delta T = 0,\pm1$  (but not $J=0 \rightarrow 0$ or $T=0 \rightarrow 0$). Fermi transitions are generated by the vector or scalar hadronic currents ($\bar\psi_p \gamma^{\mu} \psi_n$ or $\bar\psi_p \psi_n$), and they do not flip the nuclear spin. By contrast, Gamow-Teller transitions are generated by the axial-vector or tensor hadronic currents ($i\bar\psi_p \gamma^{\mu}\gamma^{5} \psi_n$ or $\bar\psi_p \sigma^{\mu \nu}\psi_n /\sqrt{2} $), and they can flip the nuclear spin. Some transitions like the neutron $\beta$ decay can have both components. Derivations of these selection rules are described in Appendix~\ref{sec:appendix4} for advanced readers.

\section{Fierz Interference}
\label{sec:Fierz}

The differential decay rate is proportional to the product of the transition matrix element $\langle f|H_{int}|i \rangle$ and its hermitian conjugate $\langle i|H^{*}_{int}|f \rangle$, where $|i\rangle$ and $|f\rangle$ are the initial and final states. Using the decomposition of $H_{int}$ into $H^{ST}_{int}$ and $H^{VA}_{int}$, there are two cross terms in the differential decay rate, $\langle f|H^{VA}_{int}|i \rangle \langle i|H^{ST*}_{int}|f \rangle$ and  $\langle f|H^{ST}_{int}|i \rangle \langle i|H^{VA*}_{int}|f \rangle$, which are also called interference terms. If both vector (axial-vector) and scalar (tensor) currents existed this interference effect should be present. To fix ideas we first consider Fermi transitions, so that only vector and scalar currents contribute. We also assume neutrinos are massless so the antineutrino final state with definite helicity has only one chirality component. Therefore, after substituting Eq.~(\ref{eq:lrva}) and Eq.~(\ref{eq:lrst}) into the interference terms, any product which involves neutrino spinors with different chiralities vanishes. Then, the non-vanishing terms are:
\begin{eqnarray}
\label{eq:Fierz1}(C_V+C_V^\prime)(C_S+ C_S^\prime)\;\langle f_l| \bar\psi_e^L \gamma_\mu \psi_\nu^L|i_l\rangle \langle i_l|\bar\psi_\nu^L\psi_e^R|f_l\rangle\;\;\\
\times\langle f_h| \bar\psi_p \gamma_\mu \psi_n|i_h\rangle \langle i_h|\bar\psi_n\psi_p|f_h\rangle,\nonumber\\
\label{eq:Fierz2} (C_V-C_V^\prime)(C_S- C_S^\prime)\;\langle f_l| \bar\psi_e^R \gamma_\mu \psi_\nu^R|i_l\rangle \langle i_l|\bar\psi_\nu^R\psi_e^L|f_l\rangle\;\;\\
\times\langle f_h| \bar\psi_p \gamma_\mu \psi_n|i_h\rangle \langle i_h|\bar\psi_n\psi_p|f_h\rangle,\nonumber
\end{eqnarray}
and their hermitian conjugates. $|i_l\rangle$ and $|f_l\rangle$ are the initial and final states of the leptons, and $|i_h\rangle$ and $|f_h\rangle$ are the initial and final states of the nucleus. Before completing the calculation of Eq.~(\ref{eq:Fierz1}) and Eq.~(\ref{eq:Fierz2}), one should pay attention to the chiral properties of these two formulae. Suppose the final state $|f_l\rangle$ involves an electron with positive helicity. In Eq.~(\ref{eq:Fierz1}), $\bar\psi_e^L$ projects out the left-handed amplitude $\sqrt{1-p_e/E_e}$ while $\psi_e^R$ projects out the right-handed amplitude $\sqrt{1+p_e/E_e}$, so Eq.~(\ref{eq:Fierz1}) is proportional to
\begin{eqnarray}
\sqrt{1+p_e/E_e}\sqrt{1-p_e/E_e}=\sqrt{1-\left(\frac{p_e}{E_e}\right)^2}=\frac{m_e}{E_e}.~~~
\end{eqnarray}
Similarly, for a final state $|f_l\rangle$ with a negative helicity electron, Eq.~(\ref{eq:Fierz1}) is still proportional to $ m_e/E_e$. So is  Eq.~(\ref{eq:Fierz2}). In fact, except for the coupling-constant factors ($C_V$, $C_V^\prime$, $C_S$ and $C_S^\prime$), Eq.~(\ref{eq:Fierz1}) and Eq.~(\ref{eq:Fierz2}) are the same,\footnote{This can be proved by explicit calculations using the spinor expressions in Appendix~\ref{sec:appendix1} but this proof will be challenging for students with no experience on the Dirac Equation or calculations with spinors. However, it is not required to understand that the interference term is proportional to $ m_e/E_e$. } after summing over the two electron helicity states. The sum of Eq.~(\ref{eq:Fierz1}) and Eq.~(\ref{eq:Fierz2}) accounts for the interference effect between vector and scalar components, called the {\em Fierz interference}\cite{fi:37} with value proportional to:
\begin{eqnarray}
\left(C_V\;C_S+C_V^\prime C_S^\prime \right)
\frac{m_e}{E_e}.
\end{eqnarray}

The $m/E$ factor is typical of situations like the present one, where there is a ``helicity mismatch''. A similar situation occurs for the highly sought-after neutrinoless double $\beta$ decay:\cite{mohapatra:04} if the neutrino is a Majorana particle\cite{pal:10} it can annihilate itself and the rate depends on a factor $m_\nu/E_\nu$. Another important example is the suppression of the decay of the negatively charged pion into an electron and its antineutrino compared to the decay into a muon and its antineutrino. The pion has zero spin so the spin of the lepton and antilepton have to be in opposite directions and momentum conservation requires them to come in opposite directions as well, so both leptons are forced into the same helicity state. The left-handedness of the weak interaction only allows positive helicity massless antineutrinos, but hinders the positive helicity massive negative-charged leptons, so the decay is suppressed by the $m/E$ factor, where $m$ is the mass of the negative-charged lepton. This factor is very small for the electron while for the muon it is of order unity and the decay proceeds mainly by $\pi^- \rightarrow \mu^- {\bar \nu_\mu}$. The decay of the positively charged pion is similar.   

For GT transitions the arguments above lead to a Fierz interference term:
\begin{eqnarray}
\left(C_A\; C_T+C_A^\prime C_T^\prime \right)
\frac{m_e}{E_e}.
\end{eqnarray}
Because of the $1/E_e$ dependence these interference effects can be identified by measuring the electron energy distributions. In the 1950's measurements had already determined that these contributions had to be small so that Fermi transitions were known to be driven by either $S$ or $V$ currents, but not by both, while GT transitions had to be driven by either $A$ or $T$ currents, but not by both.

Of course, given that antineutrinos are not really massless, there are, strictly speaking, similar terms proportional to $m_{\bar \nu}/E_{\bar \nu}$, but in practice they are negligible.

\section{$e-\nu$ Correlation}
\label{sec:enu correlation}
Consider the directional correlation between the electron and the antineutrino in $\beta$ decays from non-oriented nuclei.  Again, we classify transitions into non-spin-flip transitions (no nucleon spin flipping, $\Delta M=0$) and spin-flip transitions (some nucleon will have its spin flipped, $\Delta M=\pm 1$). We start considering non-spin-flip transitions. Because the nucleus is not oriented, we are free to choose the $+z$ direction along the momentum of the antineutrino. We first consider left-handed antineutrinos. In the $m_{\nu}=0$ limit, left-handed antineutrinos have well defined positive helicity,\footnote{We emphasize again this ``unnatural'' relationship between chirality and helicity for antiparticles. Explanations can be found in Appendix~\ref{sec:appendix1}.} and thus they are in the $|+\rangle$ spin state along the $z$ axis. In non-spin-flip transitions, the two spin projections of the two leptons are opposite to each other, so the electron is in the $|-\rangle$ state. If the weak current is vector or axial-vector, according to Eq.~(\ref{eq:lrva}) the emitted electron is also left-handed. We can repeat the chirality arguments of Sect.~\ref{sec:helicity} and determine the angular distribution of the emitted electron:
\begin{alignat}{1}
\label{eq:enucorrelationVA}
\frac{dW}{d\Omega}=\frac{1+ \left(p_e/E_e\right)\;\cos{\theta}}{4\pi}=\frac{1}{4\pi}
\left(
1 + \frac{\bm p_e}{E_e}  
\cdot  \frac{\bm p_{\bar \nu}}{E_{\bar \nu}}
\right),
\end{alignat}
where $\theta$ is the angle between ${\bm p_e}$ and ${\bm p_{\bar \nu}}$. If the weak current is scalar or tensor, according to Eq.~(\ref{eq:lrst}) the left-handed antineutrino is coupled to the right-handed electron. Therefore, the angular distribution of the emitted electron is 
\begin{alignat}{1}
\label{eq:enucorrelationST}
\frac{dW}{d\Omega}=\frac{1- \left(p_e/E_e\right)\;\cos{\theta}}{4\pi}=\frac{1}{4\pi}
\left(
1 - \frac{\bm p_e}{E_e}  
\cdot  \frac{\bm p_{\bar \nu}}{E_{\bar \nu}}
\right).
\end{alignat}
The term $\frac{\bm p_e}{E_e}  \cdot  \frac{\bm p_{\bar \nu}}{E_{\bar \nu}}$ is called the {\em $e-\nu$ correlation}, and by convention $dW/d\Omega$  is written as
\begin{alignat}{1}
\label{eq:enucorrelation}
\frac{dW}{d\Omega}=\frac{1}{4\pi}
\left(
1 + a\frac{\bm p_e}{E_e}  
\cdot  \frac{\bm p_{\bar \nu}}{E_{\bar \nu}}
\right),
\end{alignat}
where $a$ is called the $e-\nu$ correlation coefficient. In summary, for non-spin-flip transitions with left-handed antineutrinos, $a=+1$ for vector and axial-vector currents, and $a=-1$ for scalar and tensor currents. Following the arguments described above, students can work on their own to calculate the values of $a$ for right-handed antineutrinos and spin-flip transitions. In short, for right-handed antineutrinos the values of $a$ are the same as those for left-handed antineutrinos, but for spin-flip transitions the signs of $a$ are opposite.

\begin{figure}[ht]
\includegraphics[width=1.6in]{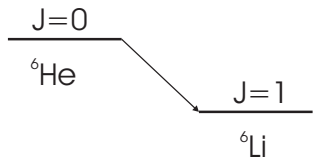}
\caption{Decay scheme for $^{6}{\rm He}$.}
\label{fig:6He-scheme}
\end{figure}

For pure Fermi transitions, where there is no angular momentum difference between the parent and daughter nucleus, the expected correlation is the same as that for non-spin-flip transitions. Therefore, for vector currents,\footnote{In pure Fermi transitions, only vector and scalar currents are involved. In pure GT transitions, only axial-vector and tensor currents are involved. Check Appendix~\ref{sec:appendix4} for further explanations.} $a=+1$, and for scalar currents $a=-1$. However, for pure GT transitions we have to consider the non-spin-flip as well as spin-flip transitions. To fix ideas we consider the case of the decay of $^{6}{\rm He}$, whose decay scheme is shown in Fig.~\ref{fig:6He-scheme}.

The daughter nucleus can have $J=1$ and $M=-1,0,+1$ so there are two spin-flip transitions and one non-spin-flip one. It can be checked that the Clebsch-Gordan coefficients corresponding to these three transitions yield the same probability for all three transitions. Thus the value of $a$ for this decay is the arithmetic average of the values of $a$ for $M=-1,0,+1$. Consequently, for axial-vector currents, $a=-1/3$, and for tensor currents $a=+1/3$. Although we focused on the decay of $^{6}{\rm He}$, it can be shown that the same result holds for any pure GT transition. Students that can maneuver comfortably with Clebsch-Gordan coefficients will figure out the way for the general proof. For the other students who are beginners we recommend trying to work out at least one other example. A summary of the values of $a$ for different cases is shown in Tab.~\ref{tab:enucorrelation}
\begin{table}
\caption{Summary of $a$ for pure Fermi and pure GT transitions for V,A currents and S,T currents.\label{tab:enucorrelation}}
\centering
\begin{ruledtabular}
\begin{tabular}{c c c}
	& Vector or axial-vector & Scalar or tensor\\
\hline
Pure Fermi & $+1$ & $-1$\\
Pure GT &$-1/3 $& $+1/3$\\
\end{tabular}
\end{ruledtabular}
\end{table}

A series of experiments were carried out in the 1950's to determine the $e-\nu$ correlation in nuclear $\beta$ decays. Because antineutrinos were difficult to detect, experimentalists measured the momentum of the recoiling-nucleus in coincidence with the momentum of the emitted electron and then calculated the momentum of the antineutrino. This kind of experiments are quite challenging because one needs to detect the rather low energy recoiling-nucleus, so lighter nuclei and larger energy release are preferred. The decay of $^{6}{\rm He}$ is an advantageous candidate, since $^{6}{\rm He}$ is one of the lightest $\beta$-decaying nuclei and the decay has a relatively large energy release (about 3.5 MeV). It is a pure GT transition and thus exclusively sensitive to axial-vector and tensor currents. One measurement published\cite{ru:53} in 1953 and confirmed\cite{ru:55} in 1955 seemed to have clearly pinned down the weak interaction in a pure GT transition to be of tensor type. In addition, measurements on $^{19}{\rm Ne}$ confirmed\cite{max:55} that the interaction was of the scalar and tensor type. Later experiments used a slightly different technique of precisely measuring the energy distribution of the recoiling-nucleus without detecting the electron. If the electron and antineutrino are preferentially emitted in the same direction (i.e. $a > 0$) the energy of the recoiling-nucleus will tend to be larger than if the leptons are preferentially emitted in opposite directions (i.e. $a < 0$). As such the shape of the recoil-nucleus energy spectrum can be used to determine $a$. A few years after the $^{19}{\rm Ne}$ experiment was published contradictory evidence from other experiments began to build a compelling case for the weak interaction being dominated by vector and axial-vector currents, not the scalar and tensor currents suggested by early experiments. Feynman gave an interesting and amusing account of these times in an article called ``{\em The 7 percent solution}''.\cite{fe:85} Eventually there was a determination of the helicity of neutrinos using a very ingenious idea;\cite{go:58} the result showed that neutrinos are predominantly left handed.  Soon after, two different groups\cite{al:59,jo:63} published new determinations of the electron-neutrino correlation with more careful measurements of the recoil-ion energy spectra from various $\beta$ emitters, definitively revealing the vector and axial-vector nature of the weak interaction. In retrospect it is not hard to see the difficulties in the first experiments and guess that more should have been demanded, but for a few years the world of physics believed weak interactions were mediated by scalar and tensor currents.

\section{$\beta$-decay Correlations in contemporary experiments}
\label{sec:ContemporaryExp}

It is clear that the weak currents are primarily of vector and axial-vector type. Nevertheless, scalar and tensor currents should not be considered as completely strange objects. In order to explain facts the SM cannot explain, such as the origin of the left-handedness of the weak interaction or the matter-antimatter asymmetry and dark matter, many theories beyond the SM have been developed. Some of these theories, for example the lepto-quark model\cite{Buchmuller1987442} and super-symmetry models,\cite{Profumo075017} do predict the existence of scalar and tensor type weak currents. To this day the most stringent limits on the coupling constants of scalar and tensor currents ( $C_S$, $C_S^\prime$, $C_T$ and $C_T^\prime$) relative to the coupling constants of vector and axial-vector currents are around a few percent.\cite{Wauters2014} Higher precision future experiments may find small but non-zero scalar or tensor currents and thus lead to a deeper understanding of the new physics beyond the SM.  A review of high precision $\beta$-decay correlation experiments was given by Severijns et al. in Ref.~\onlinecite{severijns:06}.

Higher precision experiments of the $\beta$-asymmetry and other correlations in nuclear and neutron decays have recently been published. Mund et al.\cite{Perkeo2} and Mendenhall et al.\cite{UCN:12} measured the $\beta$-asymmetry from neutron decay using cold and ultracold neutron sources. Taking advantage of technological developments that allow for faster counting and better control of systematic uncertainties a new generation of precision determination of correlations in beta-decay experiments is presently taking place or being developed for running in the near future. Examples are experiments with neutrons making ingenious use of magnetic fields\cite{Abele09,acorn,Nab,perc} and experiments with radioactive nuclei in either ion traps\cite{Flechard_2011,PRL.115.182501} or neutral atom laser traps.\cite{go:05,Vetter_2008,Behr_2009,Leredde2015}

On the other hand, experiments at the Large Hadron Collider (LHC) are also searching for scalar and tensor type interactions. A comparison of the sensitivities of low energy $\beta$-decay experiments and high-energy experiments is presented by Cirigliano et al. in Ref.~\onlinecite{Cirigliano201393,Vos15}.

\section{Conclusions}
We have presented simple derivations of nuclear $\beta$-decay correlations that originate in the selected sensitivity of the charged weak interactions for left-handed particles. In the process we described some of the history of the discoveries that led to understanding the weak interaction. In our experience students develop intuition about the properties of chirality and helicity by working through these arguments. Alternative interesting questions to consider are, for example, how the answers given here vary if one considers positron emission instead of electron emission, or how the calculations can be generalized to a mixed transition, such as neutron $\beta$ decay. The arguments used can also be applied to predict the angular distribution expected for Mott scattering, neutrino-nucleus scattering and to neutrino-electron conversion in a nuclear target among many other. 

\section{Acknowledgments}
We acknowledge the support of the U.S. Department of Energy under Grant No. DE-FG02-97ER41020. We thank Steve Ellis, Jerry Miller, Ann Nelson, Derek Storm  for helpful comments.

\appendix

\chapter{}
\section{The Dirac Equation}\label{sec:appendix1}

In order to reconcile quantum mechanics with special relativity, Dirac wrote down a Hamiltonian that is linear in ${\bm \nabla}$ for free fermions:
\begin{eqnarray}
\label{eq:Dirac_H}
H=-i{\bm{\alpha\cdot {\bm \nabla}}}+\beta m
\end{eqnarray}
where $\alpha^i=\gamma^0\gamma^i~(i=1,2,3)$ and $\beta=\gamma^0$ are $4\times4$ matrices, and the $\gamma$ matrices are
\begin{eqnarray}
\gamma^0=\left(
\begin{array}{cc}
0 & I\\
I & 0
\end{array}
\right),
\gamma^i=\left(
\begin{array}{cc}
0 & \sigma_i\\
-\sigma_i& 0
\end{array}
\right),
\gamma^5=\left(
\begin{array}{cc}
I & 0\\
0 & -I
\end{array}
\right).\nonumber\\
\end{eqnarray}
Here $I$ stands for the $2\times2$ identity matrix and $\sigma_i$ for the Pauli matrices. The matrices are such that squaring Eq.~(\ref{eq:Dirac_H}) yields a relation between momentum and energy consistent with relativity, $E^2=p^2+m^2$. We choose the Weyl representation to make the discussion on helicity and chirality easier. This Hamiltonian leads to the Dirac Equation:
\begin{eqnarray}
(i\gamma^\mu\partial_\mu-m)\psi=0
\end{eqnarray}
which describes the motion of a free fermion with a 4-component Dirac spinor $\psi$. 

The Dirac equation has two particle solutions $u_{\uparrow\downarrow}(E,{\bm{p}})e^{-iEt+i{\bm{p\cdot x}}}$ and two antiparticle solutions $v_{\uparrow\downarrow}(E,{\bm{p}})e^{iEt-i{\bm{p\cdot x}}}$. Here
\begin{eqnarray}
u_{\uparrow\downarrow}=\left[
\begin{array}{c}
\sqrt{E \mp p}\chi^{\uparrow\downarrow}\\
\sqrt{E \pm p}\chi^{\uparrow\downarrow}
\end{array}
\right],
v_{\downarrow\uparrow}=\left[
\begin{array}{c}
-\sqrt{E \mp p}\chi^{\uparrow\downarrow}\\
\sqrt{E \pm p}\chi^{\uparrow\downarrow}
\end{array}
\right],
\label{eq:Dirac Solutions}
\end{eqnarray}
where
\begin{eqnarray}
\chi^{\uparrow} =
\left[
\begin{array}{c}
1\\
0
\end{array}
\right],~~~
\chi^{\downarrow} =
\left[
\begin{array}{c}
0\\
1
\end{array}
\right]
\end{eqnarray}
correspond to the two directions of spin.
Here we suppose that $z$ is the quantization axis and momentum is also in the $+z$ direction. For antiparticle states, one can interpret $v_{\downarrow}(E,{\bm{p}})e^{iEt-i{\bm{p\cdot x}}}$ as the {\em absence} of a particle with energy $-E$, momentum $- \bm{p}$, spin along the $+z$ direction and thus negative helicity, or the {\em presence} of an antiparticle with energy $E$, momentum $\bm{p}$ and spin along the $-z$ direction and thus also negative helicity. The projectors $P_{L/R}$ in Eq.~(\ref{projectors}) project out the upper/lower two components of the spinor. Therefore, the upper/lower components have left-handed/right-handed chiralities and the amplitudes in Eq.~(\ref{eq:chirality_projection}) can be calculated. One should note that the chiralities are the eigenvalues of $\gamma^5$: $+1$ for left-handed chirality and $-1$ for right-handed chirality according to the convention used in this paper. For massive particles the chirality eigenstates are not solutions to the Dirac equation, so free massive particles cannot have a well-defined chirality. For massless particles, the lower two components of the spinor of a {\em negative}-helicity state $u_{\downarrow}$ are zero, so a state with {\em negative} helicity is equivalent to that with {\em left-handed} chirality. For massless antipaticles, the upper two components of $v_{\downarrow}$ are zero, and thus {\em negative} helicity corresponds to {\em right-handed} chirality. The complete relationships between chirality and helicity for massless particles and antiparticles are listed in Tab.~\ref{ChiralityHelicityTab}.

\begin{table}[h]
\caption{Relationships between chirality and helicity for {\em massless} particles and antiparticles.\label{ChiralityHelicityTab}}
\begin{ruledtabular}
\begin{tabular}{c  c c}
&Chirality&Helicity\\
\hline
particle&left-handed&$-$\\
particle&right-handed&$+$\\
\hline
antiparticle&left-handed&$+$\\
antiparticle&right-handed&$-$\\
\end{tabular}
\end{ruledtabular}
\end{table}

\chapter{}
\section{Isospin}\label{sec:appendix3}

In 1932 Heisenberg noted that the nuclear force between nucleons seemed to be independent on whether the nucleons were neutrons or protons. Although we now know that the strong force does have a component that depends on charge (see for example Ref.\cite{henley-garcia:2013}) this component is very small. Thus, one can approximately treat the nucleons as identical particles, except that protons and neutrons are fermions and the Pauli exclusion principle has to be satisfied within each species. Heisenberg realized that the problem is similar to that of having identical fermions with two possible spin orientations. Just like an electron with spin up is allowed in the same quantum orbits as an electron with spin down, a proton is allowed in the same quantum orbits as a neutron. As an analogy to the spin-up and spin-down states of a spin-$\frac{1}{2}$ fermion, the neutron and proton can be considered as two states of the nucleon corresponding to different {\em isospin} projections:
\begin{eqnarray}
|n\rangle &\equiv& \left| t=\frac{1}{2},t_z=-\frac{1}{2}\rangle \right.\nonumber \\
|p\rangle &\equiv& \left| t=\frac{1}{2}, t_z=\frac{1}{2}\rangle \right.\nonumber
\end{eqnarray}
which are eigenstates of the isospin operator $\hat{t}$ and isospin-projection operator $\hat{t}_z$ with eigenvalues $t$ and $t_z$. Similar to angular momentum, one can define isospin raising and lowering operators $\hat{t}^{\pm}$ and they relate the $|n\rangle$ and $|p\rangle$ as
\begin{eqnarray}
\hat{t}^{+}|n\rangle &=& |p\rangle,~ \hat{t}^{+}|p\rangle = 0\\
\hat{t}^{-}|p\rangle &=& |n\rangle,~  \hat{t}^{-}|n\rangle = 0.
\end{eqnarray}
The $\hat{t}^{+(-)}$ operator annihilates a neutron(proton) and creates a proton(neutron) with the same quantum numbers except the isospin projection. Therefore, the operators corresponding to the charged weak interactions are also represented using the $\hat{t}^{\pm}$ operators  summed over the nucleons in the nucleus.

For a given nucleus the isospin projection $T_z$ can easily be obtained as $T_z=(Z-N)/2$. The quantum number for the total isospin number $T$ is restricted to $T \ge |T_z|$. However, for nuclei close to stability, like $^{60}{\rm Co}$ and $^{6}{\rm He}$ a general useful rule is that the lowest energy states are dominated by the lowest isospin values. Thus, for example: $T(^{60}{\rm Co}) = 3$ and  $T(^{6}{\rm He})=1$.

\chapter{}
\section{Hadronic Currents And Selection Rules}\label{sec:appendix4}

The chiral properties of the lepton currents in Eq.~(\ref{eq:LYHamiltonian}), Eq.~(\ref{eq:lrva}) and Eq.~(\ref{eq:lrst}) are discussed in Sect.~\ref{sec:currents}, and the following text describes a simplification of the hadronic currents $\bar\psi_p O^i \psi_n$ in these equations. In nuclear matter, the average kinetic energy of a nucleon is less than tens of MeV, which is small compared to its rest mass ($\sim$1~GeV). In such case, the hadronic currents (after acting on initial and final states) $\langle f|\bar\psi_p O^i \psi_n|i \rangle$ can be simplified as:
\begin{eqnarray}
\label{nr_op}
\bar u_{f} \hat{t}^{+} u_{i}        &\rightarrow& \chi^{\dagger}_{f} \hat{t}^{+} \chi_{i} \nonumber \\
\bar u_{f} \hat{t}^{+}\gamma_\mu u_{i}        &\rightarrow& \chi^{\dagger}_{f}(1,{\cal O}({\bm v}/c)) \hat{t}^{+} \chi_{i} \nonumber \\
\bar u_{f} \hat{t}^{+} \gamma_\mu \gamma^5  u_{i}&\rightarrow& \chi^{\dagger}_{f}({\cal O}(v/c),{\bm \sigma}) \hat{t}^{+}  \chi_{i}\\
\bar u_{f} \hat{t}^{+} \sigma_{\mu\nu}  u_{i}&\rightarrow& \left\{
\begin{array}{c c}
{\cal O}({\bm v}/c)~~{\rm for}~\mu=0,\nu=1,2,3\\
-\epsilon_{\mu\nu\rho} \chi^{\dagger}_{f}\sigma_\rho \hat{t}^{+}  \chi_{i} ~~{\rm for}~\mu,\nu,\rho=1,2,3
\end{array}\right.
\nonumber\\
\bar u_{f} \hat{t}^{+}\gamma^5  u_{i}        &\rightarrow& {\cal O}(v/c) \nonumber 
\end{eqnarray}
where on the left-hand side the spinor $u$'s are Dirac spinors for low-energy nucleons, while on the right-hand side the spinor $\chi$'s are the corresponding Pauli spinors and the $\bm \sigma$ matrices are Pauli matrices, and $\hat{t}^{+}$ is the isospin raising operator which represents annihilating a neutron and creating a proton with the same wave function, including the spin state. For the hadronic current, the leading order of the vector and scalar currents are identical and called Fermi current. Similarly, the leading order of the axial-vector and tensor currents are identical and called Gamow-Teller current. There is no zeroth order term in the pseudo-scalar current so it is usually very small compared to other currents and has been ignored in Eq.~(\ref{eq:lrva}) and Eq.~(\ref{eq:lrst}). When calculating the decay matrix elements for an $N$-nucleon nucleus, one should use $\sum_{j}^N \hat{t}^{+}_j$, which is the total isospin raising operator $\hat{T}^{+}$, as the effective operator for the Fermi current, and $\sum_{j}^N \sigma_{j} \hat{t}^{+}_j$, which is a generic rank-1 tensor in both spin and isospin space, as the effective operator for the Gamow-Teller current. Knowing these properties of the Fermi and Gamow-Teller currents, students can use the Wigner-Eckart theorem to derive the spin and isospin selection rules mentioned at the end of Sect.~\ref{sec:currents}.

\newcommand{\noopsort}[1]{} \newcommand{\printfirst}[2]{#1}
  \newcommand{\singleletter}[1]{#1} \newcommand{\switchargs}[2]{#2#1}


\newpage   

\end{document}